# Inverse Borrmann effect in photonic crystals

A.P. Vinogradov,[a*] Yu. E. Lozovik,[b] A. M. Merzlikin,[a] A. V. Dorofeenko,[a] I. Vitebskiy,[c] A. Figotin,[c] A. B. Granovsky,[d] and A. A. Lisyansky[e]

[a] Institute of Theoretical and Applied Electromagnetism (ITAE), RAS, Izhorskaya st. 13, Moscow, 125412, Russia

[b] Institute of Spectroscopy, RAS, Moscow Region, Troitsk 142190, Russia

[c] Department of Mathematics, University of California at Irvine, Irvine, CA 92697, USA

[d] Faculty of Physics, Moscow State University, Moscow 119992, Russia

[e] Physics Department, Queens College of the City University of New York, Flushing 11367, NY, USA

## Abstract

The Borrmann effect, which is related to the microscopic distribution of the electromagnetic field inside the primitive cell, is studied in photonic and magnetophotonic crystals. This effect, well-known in x-ray spectroscopy, is responsible for the enhancement or suppression of various linear and nonlinear optical effects when the incidence angle and/or the frequency change. It is shown that by design of the primitive cell this effect can be suppressed and even inverted.

## I. Introduction

Photonic and magnetophotonic crystals (PCs and MPCs) have attracted great interest due to their unique electromagnetic properties.[1] Almost all applications of PCs are related to the existence of photonic band-gaps (BGs) where the imaginary part of Bloch wave numbers arises. This allows for the design of high-Q microcavities and optical waveguides[2] employing the Bloch wave behavior on macroscopic length-scales. Another method of the light control using PC employs varying the field distribution in a PC primitive cell. Recently, in Ref. 3, it was theoretically shown that the spatial distribution of the electric field in a MPC strongly depends on both the field frequency and the material parameters of magnetic and non-magnetic layers. As a result, a possibility of a substantial enhancement of the Faraday effect as well as nonlinear magneto-optical effects in MPCs was predicted[3-5] and experimentally verified.[4] The dependence of optical and magneto-optical properties of PCs on the spatial distribution of electromagnetic waves inside PCs can be considered as an optical analog of the Borrmann effect, which manifests itself in anomalous absorption/transmission of x-rays in near-perfect crystals.[6]

---

[*] FAX +7 (495) 484 26 33, E-mail: a-vinogr@yandex.ru

In this paper, we present a detailed analysis of the optical Borrmann effect in one-dimensional (1D) PCs. We predict that the Borrmann effect can be suppressed and even inverted by a proper design of the primitive cell.

## II. Direct and inverse Borrmann effects in 1D PCs

The electromagnetic Bloch wave propagating in a PC can be represented as

$$\vec{E}_{\vec{k}_B}(\vec{r},t) = \exp(i\vec{k}_B \cdot \vec{r} - ik_0 ct) \sum_{\{m\}} \vec{A}_{\{m\}\vec{k}_B} \exp(i\vec{G}_{\{m\}} \cdot \vec{r}), \qquad (1)$$

where, $k_0 = \omega/c$ is the renormalized frequency that is equal to the free space wave number, $c$ is the speed of light, $\vec{k}_B$ is a Bloch vector, $\vec{G}_{\{m\}}$ is a set of vectors of the reciprocal lattice, and $\vec{A}_{\{m\}\vec{k}_B}$ are harmonic amplitudes, so that

$$\vec{f}(\vec{r}) = \sum_{\{m\}} \vec{A}_{\{m\}\vec{k}_B} \exp(i\vec{G}_{\{m\}}\vec{r})$$

is a periodic function having periodicity of the PC.

We consider a 1D PC which primitive cell consists of two layers with thicknesses $d_1$ and $d_2$. Below we use $k_0$ instead of the frequency $\omega$. Particularly, in some of the plots we use dimensionless units, $k_0(d_1+d_2)$ for the frequency and $k_B(d_1+d_2)$ for the Bloch wave number, as well as $z/(d_1+d_2)$ for the space coordinate.

It turns out that depending on the frequency, $f(\vec{r})$ is increased in some layers and decreased in others. This effect is known in x-ray spectroscopy as the Borrmann effect[6] and is more pronounced at the edges of the BGs. There are two equally valid approaches for studying the edges of BGs: (*i*) one can fix the angle of incidence and vary the frequency and (*ii*) one can fix the frequency and vary the angle. The x-ray Borrmann effect corresponds to the latter approach. In this communication we consider the effect of frequency change on the field redistribution inside the bi-layer primitive cell of 1D PC. In the pass-band this redistribution can be described in terms of the periodic factor of a Bloch wave:

$$\vec{E}\cdot\vec{E}^* = \left(\vec{f}(z)e^{ik_B z}\right)\cdot\left(\vec{f}(z)e^{ik_B z}\right)^* = \vec{f}\cdot\vec{f}^*.$$

As shown below, this effect is the basis for many phenomena in PCs.

Since in the low frequency band the retardation effects in PCs are weak, we can use the homogenization theory for a qualitative explanation of the Borrmann effect. According to this theory, the effective permittivity can be defined as (see, e.g. Ref. 7 as well as Appendix A)

$$\varepsilon_{eff}\langle EE^*\rangle = \langle \varepsilon EE^*\rangle, \qquad (2)$$

where the brackets $\langle ... \rangle$ denote averaging over the volume of an elementary cell. The effective refraction index $n_{eff}$ is related to the effective permittivity in the usual manner: $n_{eff} = \sqrt{\varepsilon_{eff}}$. Although this approximation should work at low frequencies only, nevertheless, the T-matrix method calculation of the exact field distribution inside the primitive cell shows that Eq. (2) describes the behavior of the real part of the wave number above the first BG, as well (Fig. 1).[8]

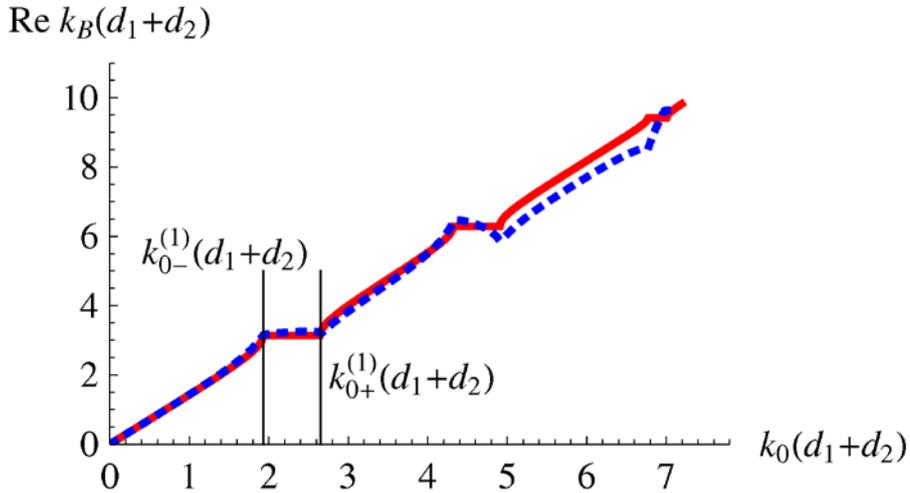

Fig. 1. Frequency dispersion of the real part of the Bloch wave number in a bi-layer PC (solid line) and of homogenized wave number $k_0(d_1+d_2)\sqrt{\langle \varepsilon EE^*\rangle/\langle EE^*\rangle}$ (dashed line). Layers with permittivities $\varepsilon_1 = 1$ and $\varepsilon_2 = 3$ have equal thicknesses $d_1 = d_2$.

The bottom $k_{0-}^{(1)}$ and top $k_{0+}^{(1)}$ frequency edges of the first band-gap correspond to the same wave number $k_B^{(1)}$.[2] Therefore, normalizing the fields at different frequencies by the condition $\langle E_-E_-^*\rangle = \langle E_+E_+^*\rangle = 1$, where the subscripts "+" and "−" denote top and bottom edges of a BG, respectively, we can write the following equality:

$$k_B^{(1)} = k_{0-}^{(1)} n_- = k_{0+}^{(1)} n_+$$

where $n_- = \sqrt{\langle \varepsilon E_-E_-^*\rangle}$ and $n_+ = \sqrt{\langle \varepsilon E_+E_+^*\rangle}$. As a result, the inequality $k_{0-}^{(1)} < k_{0+}^{(1)}$ leads to the inequality for reflection indices: $n_- > n_+$ or

$$\langle \varepsilon E_-E_-^*\rangle > \langle \varepsilon E_+E_+^*\rangle.$$

Taking into account that due to the normalization condition, $\langle E_\gamma E_\gamma^*\rangle_l = 1 - \langle E_\gamma E_\gamma^*\rangle_h$ ($\gamma$ stands for "+" or "−", $h$ and $l$ denote higher- and lower-permittivity layers), and that

$$\langle \varepsilon E_-E_-^*\rangle = \varepsilon_h \langle E_-E_-^*\rangle_h + \varepsilon_l \langle E_-E_-^*\rangle_l,$$

$$\langle \varepsilon E_+E_+^*\rangle = \varepsilon_h \langle E_+E_+^*\rangle_h + \varepsilon_l \langle E_+E_+^*\rangle_l,$$

we arrive at the following inequalities:

$$\varepsilon_h \langle E_-E_-^*\rangle_h > \varepsilon_h \langle E_+E_+^*\rangle_h, \qquad (3a)$$

$$\varepsilon_l \langle E_-E_-^*\rangle_l < \varepsilon_l \langle E_+E_+^*\rangle_l. \qquad (3b)$$

It means that the field and the energy concentration in the high permittivity layer at the bottom of the BG is greater than that at the top of the BG whereas the field and the energy concentration in the lower permittivity layer is greater at the top of the BG than that at the bottom of the BG, as shown in Fig. 2. This effect can be considered as the frequency analog of the Borrman effect.[2,6]

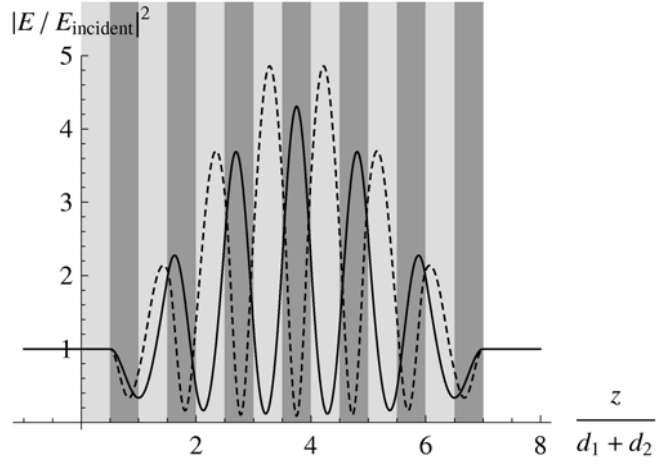

Fig. 2. The results of the exact solution of the 1D Maxwell equation: the electric field $|E|^2$ distribution in a PC slab at frequencies of the nearest to BG edges Fabri-Perot resonances: the solid and the dashed lines present the case below ( $k_0(d_1+d_2)=1.55$ ) and above ( $k_0(d_1+d_2)=2.77$ ) the first BG, respectively. Darker and lighter areas depict layers with higher, $\varepsilon_2=3$, and lower, $\varepsilon_1=1$, permittivities, $d_1=d_2$. It is assumed that the slab is in vacuum, $\varepsilon_{ext}=1$.

The change in the energy distribution happens both inside the BG and inside the adjacent pass-bands (this effect can be observed even in the case of a low contrast PC, see the results of the exact solution of the Maxwell equations by the transfer matrix method shown in Fig. 3). At low frequency, $k_0 \ll \pi/(\sqrt{\varepsilon_1}d_1+\sqrt{\varepsilon_2}d_2)$, the fraction of the energy of the electric field in the high permittivity layer $\langle \varepsilon EE^* \rangle_h / \langle \varepsilon EE^* \rangle$ is equal to the "relative electrical thickness" of this layer $\varepsilon_2 d_2/(\varepsilon_1 d_1+\varepsilon_2 d_2)$. This fraction increases with frequency growth, reaching the maximum at the bottom edge of the first BG. While passing the BG, the fraction decreases in agreement with inequality (3).

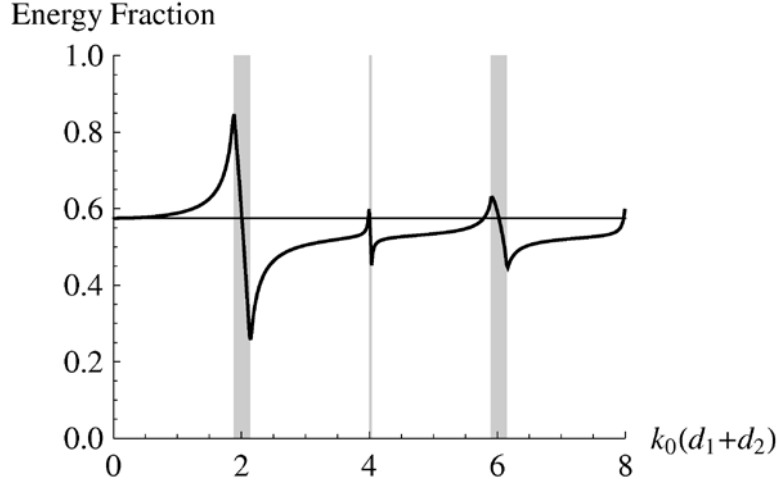

Fig. 3. The frequency dependence of the fraction of the electric field energy in the high permittivity layer for the low contrast PC with $d_1/d_2 = 1.108$, $\varepsilon_1 = 2$ and $\varepsilon_2 = 3$. (These values of thicknesses and permittivities correspond to the parameters $K_0 = 1$ and $\alpha = 0.05$, which are defined below). BG regions are shaded. The horizontal line corresponds to the uniform distribution of the field.

It can be shown (see Appendix A) that if the deviation of permittivity $\delta\varepsilon(z)$ from the average value $\varepsilon_0$ is small, the perturbation theory in $\delta\varepsilon/\varepsilon_0$ gives the same real part of the wave number as the homogenization procedure does not only for the first BG but also for higher BGs. In the linear approximation in $\delta\varepsilon/\varepsilon_0$, the energy concentration in the high permittivity layer is always greater at the bottom of the BG than that at the top of the BG (see Fig. 3). Below we refer to this situation as the direct Borrmann effect. The situation where the energy concentration in the high permittivity layer is smaller at the bottom of the BG than that at the top of the BG constitutes the inverse Borrmann effect. As far as we know, the inverse Borrmann effect was never discussed in literature.

Thus, until the linear perturbation is valid, we deal with the direct Borrmann effect that can be explained in terms of the homogenization theory. For the high contrast of permittivity ($\varepsilon_2 \gg \varepsilon_1$) both perturbation and homogenization approaches are not applicable (see Appendix A). However, as one can see in Fig. 4, the discrepancy between the values of the wavenumber obtained as solutions of the exact dispersion equation[9] and of Eq. (2) becomes significant in the vicinity of BGs only.

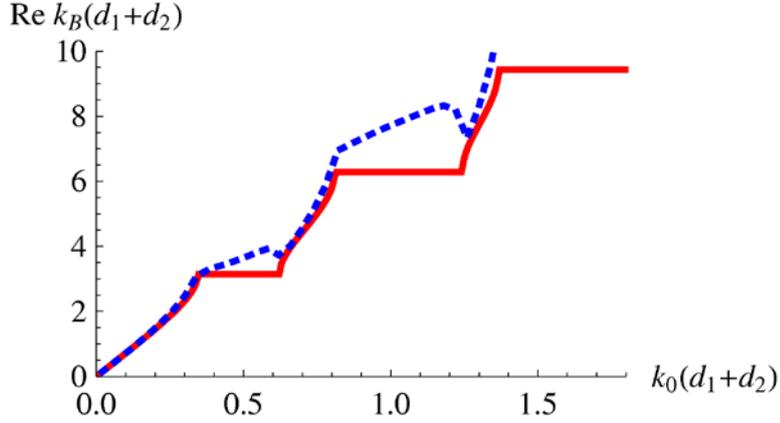

Fig. 4. Frequency dispersion of the real part of the Bloch wave number in a bi-layer PC (solid line) and of homogenized wave number $k_0(d_1+d_2)\sqrt{\langle \varepsilon EE^* \rangle / \langle EE^* \rangle}$ (dotted line). Layers with permittivity values $\varepsilon_1 = 1$ and $\varepsilon_2 = 100$ have equal thicknesses, $d_1 = d_2$.

To move further it is convenient to choose the primitive cell of the bi-layer PC as a three-layer combination in which one of the layers is sandwiched by the half-layers of the other type. The internal layer can be considered as a cavity of the Fabry-Perot resonator. Since there are two choices of the primitive cell, the system under consideration has two sets of resonances: $k_0 d_1 \sqrt{\varepsilon_1} = \pi n_1$ and $k_0 d_2 \sqrt{\varepsilon_2} = \pi n_2$. Near these frequencies the PC can be considered as a chain of the coupled Fabry-Perot resonators. In accordance with the tight-binding theory, each resonator eigenfrequency broadens into a pass-band. Thus, the properties of the Bloch waves are determined by the corresponding resonance. Taking into account that at the resonant frequencies the reflection coefficient from the central layer is equal to zero and the Bloch wave in the complementary layer reduces to a single traveling wave, one can show by direct calculation that the ratio $\langle \varepsilon EE^* \rangle_h / \langle \varepsilon EE^* \rangle_l$ is proportional to $\sqrt{\varepsilon_h / \varepsilon_l}$ near the low permittivity resonance and to $\sqrt{\varepsilon_l / \varepsilon_h}$ near the high permittivity resonance. Since $\sqrt{\varepsilon_h / \varepsilon_l} \gg \sqrt{\varepsilon_l / \varepsilon_h}$, the most part of the energy turns out to be contained in the cavity layer. We may expect significant field redistribution while passing a BG between consequent pass-bands, related to resonances in the different layers being considered as cavities of Fabry-Perot resonators.

To understand how this redistribution of the energy leads to the direct and inverse Borrmann effects it is convenient to consider PCs of different structures. If the structure of the cell changes, the band structure changes as well. To trace this dependence for a bi-layer primitive cell, it is convenient to parameterize the structure by a single parameter $\alpha$ defined by the following equations: $d_1 = \pi(1-\alpha)/(K_0\sqrt{\varepsilon_1})$ and $d_2 = \pi(1+\alpha)/(K_0\sqrt{\varepsilon_2})$, so that

$$\alpha = \frac{d_2\sqrt{\varepsilon_2} - d_1\sqrt{\varepsilon_1}}{d_2\sqrt{\varepsilon_2} + d_1\sqrt{\varepsilon_1}}, \tag{4}$$

where $K_0 = 2\pi/D$, $D$ is a dimensionless cell optical thickness. For any values of parameters $k_0$ and $\alpha$ one obtains $k_0 d_1 \sqrt{\varepsilon_1} + k_0 d_2 \sqrt{\varepsilon_2} = 2\pi k_0 / K_0 = k_0 D$. Since $D$ does not depend on $\alpha$, by varying the parameter $\alpha$ within the interval $-1 < \alpha < 1$ we change the relative thicknesses of layers keeping the value of the optical thickness fixed. The band structure of a PC is shown in Fig. 5 for the case of large contrast. In the case of low contrast, the picture of the band structure looks similar but areas corresponding to BGs are smaller.

As it shown in Fig. 5, there are points with the zero-width BGs.[10] These are the points of crossing of dotted and dashed lines, corresponding to the transparency conditions

$$k_0 d_1 \sqrt{\varepsilon_1} = \pi n_1,$$
$$k_0 d_2 \sqrt{\varepsilon_2} = \pi n_2. \tag{5}$$

At these points the ratio of the optical paths of the layers is a rational number. Thus, the condition for the formation of the *N*-th zero-width BG is

$$k_0 d_1 \sqrt{\varepsilon_1} + k_0 d_2 \sqrt{\varepsilon_2} = \pi N, \tag{6}$$

where $N = n_1 + n_2$. In this case, in spite of the fulfillment of the Bragg condition, no BG is formed because the transfer matrix of each layer is equal to the unit matrix and the PC is equivalent to a uniform medium without BGs.[10] It is worth mentioning that in the case of small contrast in the vicinity of the points of zero-width BG, the linear perturbation theory considered above becomes invalid due to vanishing of the linear term. Then high-order terms dominate, which sometimes leads to inversion to the Borrmann effect (see Appendix B).

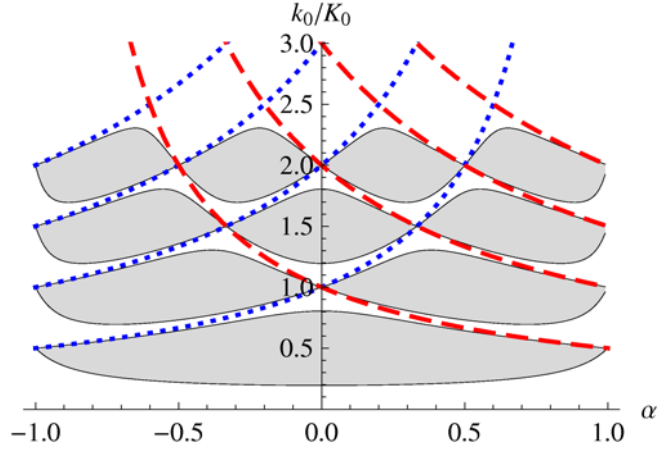

Fig. 5. Evolution of the band structure with the change of the parameter $\alpha$. Dotted ( $k_0/K_0 = n_1/(1-\alpha)$ ) and dashed ( $k_0/K_0 = n_2/(1+\alpha)$ ) curves correspond to the Fabri-Perot resonances $k_0 d_1 \sqrt{\varepsilon_1} = \pi n_1$ and $k_0 d_2 \sqrt{\varepsilon_2} = \pi n_2$, respectively. BG regions are shaded. The permittivity values of the layers are $\varepsilon_1 = 1$ and $\varepsilon_2 = 100$. The width of the second BG becomes zero at $k_0/K_0 = 1$.

In Fig. 5 the resonance conditions are satisfied at the dashed and dotted lines. At the resonant frequencies, the energy is concentrated inside layers that play the role of the Fabry-Perot cavities (Fig. 6). Crossing a point of a zero-width BG at a fixed frequency changes the order of resonances (Fig. 5) and brings us to inversion of the field distribution (see Fig. 6). Thus, we arrive into the inverse Borrmann effect. Since the first BG is always non-zero, the inverse Borrmann effect occurs for the higher BGs only. As shown in Fig. 6a, at $\alpha = 0.15$ the direct effect occurs near the first, second and forth BGs, while the inverse effect occurs near the third and fifth BGs, at $\alpha = -0.15$ (Fig. 6b) one can see the direct effect near the first, third and fifth BGs, while the inverse effect occurs near the second and forth BGs.

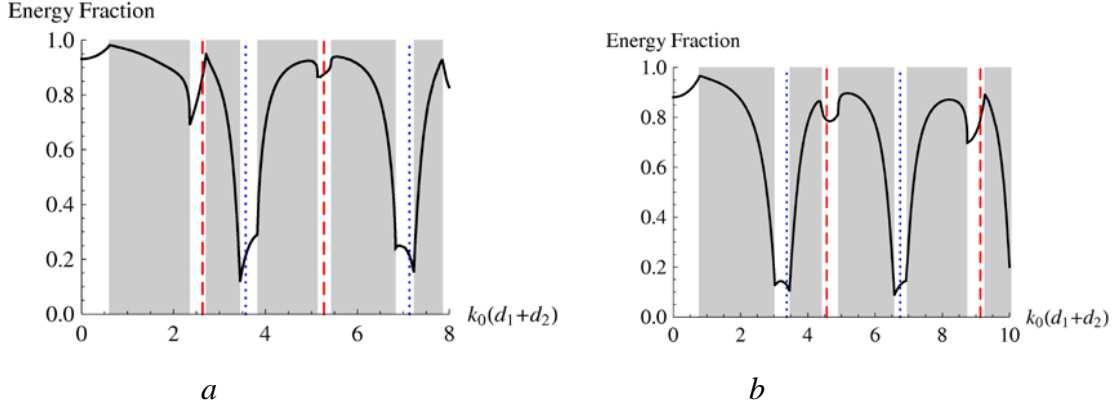

Fig. 6. The frequency dependence of the energy fraction in the high permittivity layer (solid line) for different structures of the primitive cell at a high contrast PC ( $\varepsilon_1 = 1$ and $\varepsilon_2 = 100$ ). (*a*) $\alpha = 0.15$, (*b*) $\alpha = -0.15$. BG regions are shaded. The vertical lines show frequencies of the Fabry-Perot resonances that are shown by dashed and dotted lines in Fig. 5. The direct Borrmann effect occurs when of the energy fraction decreases with the frequency inside a BG. The inversion of this dependency corresponds to the inverse Borrmann effect. The latter phenomenon is observed when a pass-band containing the Fabry-Perot resonance in the lower-permittivity layer (dotted vertical lines) is below the BG and a pass-band containing higher-permittivity layer Fabry-Perot resonance (dashed vertical lines) is above the BG.

## III. Manifestation of the Borrmann effect

One can observe the Borrmann effect through the enhancement of the Faraday rotation induced by a slab of a MPC near the Fabry-Perot resonance of the Bloch waves in a MPC slab.[1] It can be shown that in the vicinity of the resonance the angle of the Faraday rotation $\theta$ is equal to $-Qg W_{MO}/\varepsilon W$, where $Q$ is the Q-factor of the Fabry-Perot resonance, $W_{MO}$ and $W$ are the electric energies accumulated in a magnetoactive layer and in the whole primitive cell, respectively, and $g$ is the gyration constant proportional to the static magnetization. Thus, the higher the concentration of the field in magneto-optical layers, the greater the Faraday rotation. Therefore, the Faraday rotation angles should be notably different at different sides of a BG. The experimental observation and computer simulation of this effect near the first BG (850 nm) were reported in Ref. 4.

To observe the inverse Borrmann effect we should consider the second BG. To illustrate the inversion we examine a MPC sample similar to the one studied in Ref. 4.

We consider a sample 10×(SiO$_2$/Bi:YIG) with $d_1 = \pi(1-\alpha)/(K_0\sqrt{\varepsilon_1})$ and $d_2 = \pi(1+\alpha)/(K_0\sqrt{\varepsilon_2})$. At $K_0 = 2\pi/850$ nm$^{-1}$ the second BG is near the wavelength of 850 nm (Fig. 7). The zero-width BG is at $k_0 d_1 \sqrt{\varepsilon_1} = \pi$, $k_0 d_2 \sqrt{\varepsilon_2} = \pi$ ($\alpha = 0$). In Fig. 7a the MPC has $\alpha = +0.4$ that corresponds to the situation below the zero-width BG, whereas in Fig. 7b the MPC has $\alpha = -0.4$ that corresponds to the situation above the zero-width BG. As Fig. 7 shows, crossing of the zero-width BG is accompanied by the inversion of the Borrmann effect, as predicted in this paper.

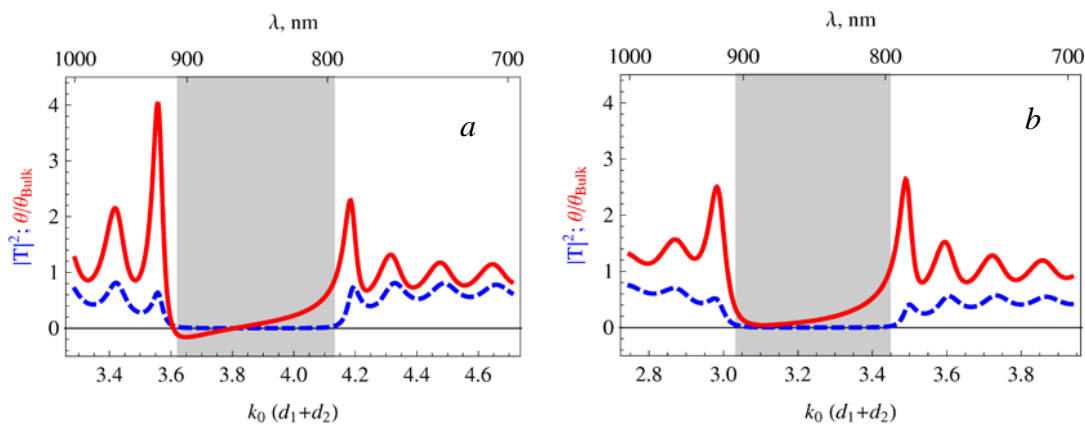

Fig. 7. The transmittance (blue line) and the normalized Faraday rotation (red line) of a slab of a MPC for frequencies (a) below and (b) above the zero-width BG. BG regions are shaded.

## IV. Conclusions

We demonstrated that by varying parameters of a PC cell it is possible to change the spatial distribution of the electromagnetic field inside the PC, and as a result, to significantly enhance or suppress linear and non-linear optical and magneto-optical effects. It was shown that there are two quite opposite cases: the direct and inverse optical Borrmann effects. In the first case, for the frequencies at the bottom of the BG, the energy concentration in a high permittivity layer is greater than that for frequencies at the top of the BG. The direct Borrmann effect was confirmed in recent experiments [4]. In the case of the inverse Borrmann effect, the spatial distribution of energy is inverted and the electric field for frequencies at the bottom of the BG is mainly localized in the low permittivity layers. The conditions for the direct and inverse Borrmann effects were found and it was shown that in the first BG the inverse Borrmann effect does not exist for any contrast of dielectric permittivities of layers.

**Acknowledgement**

This work was partly supported by RFBR grants as well as by AFSOR, Grant No. FA9550-07-1-0391.

**References**

[1] M. Inoue, K. Arai, T. Fujii, M. Abe, JAP **83**, N11, 6768 – 6770 (1998); M. Inoue, R. Fujikawa, A. Baryshev, A. Khanikaev, P. B. Lim, H. Uchida, O. A. Aktsipetrov, A. A. Fedyanin, T. V. Murzina, and A. B. Granovsky, J. Phys. D: Appl. Phys. **39**, R151(2006); M. J. Steel, M. Levy, and R. M. Osgood, Jr., J. Lightwave techn, **18**, 1297 (2000); S. Kahl, A.M. Grishin, Appl. Phys. Lett., **84**, N9, 1438 (2004).

[2] R. D. Meade, A. M. Rappe, K. D. Brommer, J. D. Joannopoulos, and O. L. Alherhand, Phys. Rev. B **48**, 8434 (1993); G.S. Johnson and D.J. Joannopoulos, Photonic Crystals. The road from Theory to Practice, Boston: Kluwer, 2002.

[3] S. Erokhin, A. Vinogradov, A. Granovsky, and M. Inoue, Physics of the Solid State, **49**, 497 (2007); A. Vinogradov, A. Granovsky, S. Erokhin, and M. Inoue Journal of Communication Technology and Electronics, **49**, 88 (2004); **49**, 682 (2004).

[4] A. B. Khanikaev, A. B. Baryshev, P. B. Lim, H. Uchida, M. Inoue, A. G. Zhdanov, A. A. Fedyanin, A. I. Maydykovskiy, and O. A. Aktsipetrov, Phys Rev B **78**, 193102 (2008).

[5] A. Figotin and I. Vitebskiy, Phys. Rev. B, **77,** 104421 (2008); S. G. Erokhin, A. A. Lisyansky, A. M. Merzlikin, A. P. Vinogradov, and A. B. Granovsky, Phys Rev B, **77,** 233102 (2008).

[6] G. Borrmann, Physik Z., **42**, 157 (1941); B. W. Battermen and H. Cole, Rev. Mod. Phys. **36**, 681 (1964); Z. Zhang and S. Satpathy, Phys. Rev. Lett. **65**, 2650 (1990).

[7] E. P. Velikhov and A. M. Dykhne, Proc. VI Int. Symp. Ion. Phen. In Gases, Paris p. 511 (1963); D. J. Bergman, Physics Rep. **43**, 377 (1978); A. P. Vinogradov, Electrodynamics of composite materials Moscow: URSS (2002) p. 38 (in Russian).

[8] The theory gives pure real values of the wave number and cannot describe the effects of the band-gap, which are due to the imaginary part of the wave number.

[9] S. M. Rytov, Akust. Zh. **2**, 71 (1956) [Sov. Phys.–Acoust. **2**, 68 (1956)]; L. M. Brekhovskikh, Waves in Layered Media (Academic Press, NY, 1960)

[10]I. Nusinsky and A. A. Hardy, Phys Rev B **73**, 125104 (2006); M. de Dios Leyva and J. Lopegzo Gondar, phys. stat. sol. (b) **128,** 575 (1985); V. Milanovic and D. Tjapkin, phys. stat. sol. (b) **110,** 687 (1982); J. Zak Phys Rev B **32**, 2218 (1985).

## Appendix A

Let us consider a homogeneous medium with permittivity $\varepsilon_0$. If the permittivity distribution is periodically modulated with $\delta\varepsilon(z)$, we obtain a PC with the permittivity distribution $\varepsilon(z) = \varepsilon_0 + \delta\varepsilon(z)$. In accordance with the band theory, the *n*-th BG is formed at the wave number $k_B^{(n)} = \pi n / d$ (*d* is a period of the PC). In a homogeneous medium this wave number corresponds to the normalized frequency $k_0^{(n)} = k_B^{(n)}/\sqrt{\varepsilon_0} = \pi n /(\sqrt{\varepsilon_0} d)$. Assuming that $|\delta\varepsilon|/\varepsilon_0 \ll 1$ and employing the perturbation theory we can find frequencies $k_{0\pm}^{(n)}$, where subscripts "+" and "−" correspond to the top and bottom BG edges, respectively:

$$\left(k_{0\pm}^{(n)}\right)^2 - \left(k_0^{(n)}\right)^2 = -\left(k_0^{(n)}\right)^2 \frac{\int \delta\varepsilon \, |E_{0\pm}^{(n)}|^2 \, dz}{\varepsilon_0 \int |E_{0\pm}^{(n)}|^2 \, dz}.$$

As we deal with the degenerated case, the nonperturbed field $E_{0\pm}^{(n)}$ should be chosen to be different for the lower and upper band edges. To fix this choice we have to take into account that at BG edges there is no energy flux, and that the fields should be orthogonal functions. For a symmetric cell, they are to be odd and even functions with respect to the center of the cell. The choice of these functions as sine and cosine having an integer number of half-periods in the cell diagonalizes the secular matrix.

Transforming the latter expression into

$$\left(k_{0\pm}^{(n)}\right)^2 \frac{\varepsilon_0 \int |E_{0\pm}^{(n)}|^2 \, dz}{\int (\varepsilon_0 - \delta\varepsilon) |E_{0\pm}^{(n)}|^2 \, dz} = \left(k_0^{(n)}\right)^2 = \frac{\left(k_B^{(n)}\right)^2}{\varepsilon_0}$$

and taking into account that $|\delta\varepsilon|/\varepsilon_0 \ll 1$ we arrive at:

$$\left(k_B^{(n)}\right)^2 = \left(k_{0\pm}^{(n)}\right)^2 \varepsilon_0 \left(1 - \frac{\int \frac{\delta\varepsilon}{\varepsilon_0} |E_{0\pm}^{(n)}|^2 \, dz}{\int |E_{0\pm}^{(n)}|^2 \, dz}\right)^{-1}$$

$$\approx \left(k_{0\pm}^{(n)}\right)^2 \varepsilon_0 \left(1 + \frac{\int \frac{\delta\varepsilon}{\varepsilon_0} |E_{0\pm}^{(n)}|^2 \, dz}{\int |E_{0\pm}^{(n)}|^2 \, dz}\right) = \left(k_{0\pm}^{(n)}\right)^2 \frac{\int \varepsilon |E_{0\pm}^{(n)}|^2 \, dz}{\int |E_{0\pm}^{(n)}|^2 \, dz}.$$

(A1)

Eq. (2) immediately follows from Eq. (A1): $\left(k_B^{(n)}\right)^2 = \left(k_{0\pm}^{(n)}\right)^2 \varepsilon_{eff\pm}^{(n)}$, where

$$\varepsilon_{eff\pm}^{(n)} = \frac{\int \varepsilon |E_{0\pm}^{(n)}(z)|^2 \, dz}{\int |E_{0\pm}^{(n)}(z)|^2 \, dz}.$$

The function $E_{0\pm}^{(n)}(z)$ as well as the value of $\left(k_{0\pm}^{(n)}\right)^2$ is different at the lower and upper edges of the BG (this fact is shown by the symbol '$\pm$') whereas $\left(k_B^{(n)}\right)^2$ is the same. Using the homogenization theory one can predict the Borrmann effect but cannot predict its inversion.

The above arguments are valid until the BG is determined by the unperturbed frequency $k_{01}^{(n)} = \pi n / \left(\sqrt{\langle\varepsilon\rangle} d\right)$. Indeed, the BG position is determined by the quantity $k_{02}^{(n)} = \pi n / \left(\langle\sqrt{\varepsilon}\rangle d\right)$ (see Fig. 8). Thus, the high frequency limit is determined by the condition that the frequencies $k_{01}^{(n)}$ and $k_{02}^{(n)}$ belong to the $n$-th BG:

$$\left(1/\langle\sqrt{\varepsilon}\rangle - 1/\sqrt{\langle\varepsilon\rangle}\right) << \frac{\left(k_{0+}^{(n)}\right)^2 - \left(k_{0-}^{(n)}\right)^2}{\left(k_0^{(n)}\right)^2}.$$

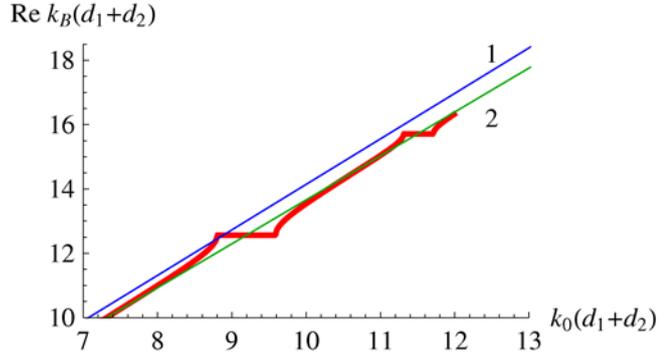

Fig. 8. Thick line is a fragment of the same dispersion curve as in Fig. 1. The straight lines are dispersion curves of nonperturbed medium at different choice of $\varepsilon_0$. Line 1 corresponds to the homogenization wave vector $k_B = k_0\sqrt{\langle\varepsilon\rangle}$, line 2 to the exact theory dependence $k_B = k_0 \langle\sqrt{\varepsilon}\rangle$.

## Appendix B

As it follows from Eqs. (4)-(6), the conditions for zero-width BGs are satisfied at a series of discrete values of $\alpha = \alpha_{zero}$ (the points of crossing of dashed and dotted lines in Fig. 5), that can be found by dividing the segment $-1 \leq \alpha \leq 1$ into $N$ equal parts. Near these points the linear perturbation theory is invalid, since the linear term becomes zero and higher-order terms should be taken into account. In these frequency domains we observe the following behavior: for $\alpha > \alpha_{zero}$ the direct Borrmann effect is realized (Fig. 3), for $\alpha < \alpha_{zero}$ but close to $\alpha_{zero}$ we observe the inverse Borrmann effect at the second BG (Fig. 9a): the energy concentration in the high permittivity layer is greater at the top of the BG than that at the bottom of the BG. Further increase of $(\alpha_{zero} - \alpha)$ leads to disappearance of the Borrmann effect (the fraction of the electric field energy is the same on both sides of the BG see Fig. 9b) and then to the direct Borrmann effect (Fig. 9c). Since the value of $\alpha$, given by Eq. (4), can be varied by changing the relative thickness of the layers for fixed values of the layers' permittivities, the inverse Borrmann effect can be observed even at low contrast. However, as one can see in Fig. 5, there is no point $\alpha_{zero}$ and corresponding to it the inverse Borrmann effect for the first BG.

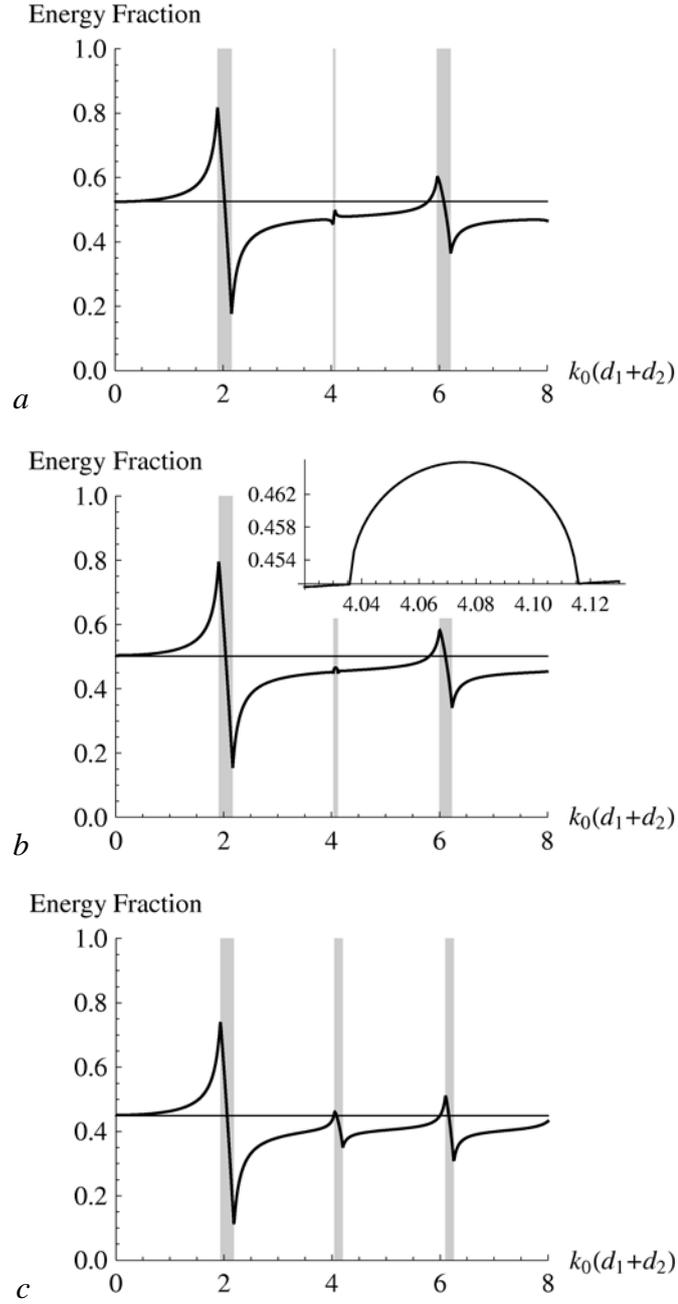

Fig. 9. The frequency dependence of the energy fraction in the high permittivity layer (solid line) for different structures of the primitive cell at a low contrast PC near zero-width BG. (*a*) $\alpha = -0.05$; (*b*) $\alpha = -0.98$, the detailed behavior near the second BG showed in the inset, confirms that the fraction of the electric field energy is the same on both sides of the BG; (*c*) $\alpha = -0.2$. $\alpha_{zero} = 0$. Other parameters are the same as in Fig. 3. BG regions are shaded.